\documentclass[10pt]{elsarticle}
\journal{Complexity}
\usepackage{amsmath}
\usepackage{indentfirst}
\usepackage{amsfonts,amssymb}
\usepackage{mathrsfs}
\usepackage{amsthm}
\usepackage{lineno,hyperref}
\modulolinenumbers[5]
\newtheorem{Definition}{Definition}[section]
\newtheorem{Theorem}{Theorem}[section]

\newtheorem{Proposition}{Proposition}[section]

\numberwithin{equation}{section}
\newtheorem{Remark}{Remark}[section]

\begin{document}
	
	\begin{frontmatter}
		
		\title{Complex risk statistics with scenario analysis}

		\author{Fei Sun\corref{}}
		\cortext[mycorrespondingauthor]{Corresponding author}
		\address{School of Mathematics and Computational Science, Wuyi University, Jiangmen 529020, China}
		\ead{fsun.sci@outlook.com}

		\author{Yichuan Dong\corref{mycorrespondingauthor}}
		\address{National Supercomputing Center in Shenzhen, Shenzhen 518055,  China}
		
		
		\ead{dongyc@nsccsz.cn}

\begin{abstract} Complex risk is a critical factor for both intelligent systems and risk management. In this paper, we consider a special class of risk statistics, named complex risk statistics. Our result provides a new approach for addressing complex risk, especially in deep neural networks. By further developing the properties related to complex risk statistics, we are able to derive dual representation for such risk. 
\end{abstract}


\end{frontmatter}

%\linenumbers

\section{Introduction}
Research on complex risk is a popular topic in both intelligent systems and theoretical research, and complex risk models have attracted considerable attention,  especially in deep neural networks. 
The quantitative calculation of risk involves
two problems: choosing an appropriate complex risk model and allocating complex risk to individual components. This has led to further research on complex risk.

In a seminal paper, Artzner et al. \cite{4,5} first introduced the class of coherent risk measures. Later Sun et al. \cite{21} and Sun and Hu \cite{22} focused on set-valued risk measures.
However, traditional risk measures may fail to describe the characteristics of complex risk. This concept has promoted the study of complex risk measures. Systemic risk measures were axiomatically introduced by Chen et al. \cite{8}. Other studies of complex risk measures include those of Acharya et al. \cite{1}, Armenti et al. \cite{3}, Biagini et al. \cite{6}, Brunnermeier and Cheridito \cite{7}, Feinstein et al. \cite{11}, Gauthier et al. \cite{13},  Tarashev et al. \cite{23}, and the references therein.

From the statistical point of view, the behaviour of a random variable can be
characterized by its observations, the samples of the random variable.  \cite{18} and \cite{20} first
introduced the class of natural risk statistics, the corresponding dual
representations are also derived. An alternative proof of the dual representation of the
natural risk statistics was also derived by \cite{2}.
Later, \cite{26} obtained dual representations for convex risk statistics,
and the corresponding results for quasiconvex risk statistics were obtained by \cite{25}.
Deng and Sun \cite{15} focused on the regulator-based risk statistics for portfolios.
However, all of these risk statistics are designed to quantify risk of simple component (i.e. a random variable) by its samples. A natural question is determining how to quantify complex risk by its samples.

The main focus of this paper is a new class of risk statistics, named complex risk statistics. In this context, we divide the measurement of complex risk into two steps. Our results illustrate that each complex risk statistic can be decomposed into a clustering function and a simple risk statistic, which provides a new approach for addressing complex risk. By further developing the axioms related to complex risk statistics, we are able to derive their dual representations.

The remainder of this paper is organized as follows. In Sect.~\ref{sec:3}, we derive the definitions related to complex risk statistics. Sect.~\ref{sec:4} discusses a new measurement of complex risk statistics. Finally, in Sect.~\ref{sec:5}, we consider the dual representations of complex risk statistics.

\section{The definition of complex risk statistics}
\label{sec:3}

In this section, we state the definitions related to complex risk statistics. 
Let $\mathbb{R}^{d}$ be the $d$-dimensional Euclidean space, $d\geq1$. 
For any $ x=(x_{1},\ldots, x_{d})$, $y=(y_{1},\ldots, y_{d})\in \mathbb{R}^{d}$, $ x\leq y $ means $ x_{i}\leq y_{i} $, $ 1\leq i\leq d $. 
For any positive integer $ k_{i} $, the element $X$ in product Euclidean space $ \mathbb{R}^{k_{1}}\times \mathbb{R}^{k_{2}}\times \ldots\times \mathbb{R}^{k_{d}} $ is denote by $X:=(X_{1}^{1},\ldots, X_{k_{1}}^{1}, X_{1}^{2},\ldots, X_{k_{2}}^{2},\ldots\ldots X_{1}^{d},\ldots, X_{k_{d}}^{d})$. For any
$X,Y\in \mathbb{R}^{k_{1}}\times \mathbb{R}^{k_{2}}\times \ldots\times \mathbb{R}^{k_{d}}$ and $ 1\leq i\leq d $,  $X\succeq Y$ means $ \sum_{j=1}^{k_{i}}X^{i}_{j}\leq \sum_{j=1}^{k_{i}}Y^{i}_{j}$.
From now on, the addition and multiplication are all defined pointwise.
$\langle X,Y   \rangle =\sum_{i=1}^{d}(\sum_{j=1}^{k_{i}}X^{i}_{j} \sum_{j=1}^{k_{i}}Y^{i}_{j})$.
$\langle x,y   \rangle =\sum_{i=1}^{d}x_{i}y_{i}$
For any
$X\in \mathbb{R}^{k_{1}}\times \mathbb{R}^{k_{2}}\times \ldots\times \mathbb{R}^{k_{d}}$, $X_{[k_{i}]}:=(0,\ldots\ldots, 0,  X_{1}^{i},\ldots, X_{k_{i}}^{i}, 0,\ldots\ldots, 0)\in 0^{k_{1}}\times \ldots  \times 0^{k_{i-1}}\times \mathbb{R}^{k_{i}}\times 0^{k_{i+1}}\times  \ldots\times 0^{k_{d}}$.

\begin{Definition}
	A simple risk statistic is a function $\varrho:$ $\mathbb{R}^{d} \rightarrow \mathbb{R}\cup \{+\infty\}$ that satisfies the following properties,
	\begin{description}
		\item[A1] Monotonicity: for any $x,y\in \mathbb{R}^{d}$, $x\geq y$ implies $\varrho(x)\geq \varrho(y)$;
		\item[A2] Convexity: for any $x,y\in \mathbb{R}^{d}$ and $\lambda\in[0,1]$, $\varrho\big(\lambda x+(1-\lambda)y\big)\leq \lambda\varrho(x)+(1-\lambda)\varrho(y)$.\\
	\end{description}
\end{Definition}

\begin{Remark}
	The properties $\mathbf{A1}-\mathbf{A2}$ are very well known and have been studied in detail in the study of risk statistics. 
\end{Remark}

\begin{Definition}
	A clustering function is a function $\phi:\mathbb{R}^{k_{1}}\times \mathbb{R}^{k_{2}}\times \ldots\times \mathbb{R}^{k_{d}} \rightarrow \mathbb{R}^{d}$  that satisfies the following properties,
	\begin{description}
		\item[B1] Monotonicity: for any $X,Y\in \mathbb{R}^{k_{1}}\times \mathbb{R}^{k_{2}}\times \ldots\times \mathbb{R}^{k_{d}}$, $X\succeq Y$ implies $\phi(X)\geq\phi(Y)$;
		\item[B2] Convexity: for any $X,Y\in \mathbb{R}^{k_{1}}\times \mathbb{R}^{k_{2}}\times \ldots\times \mathbb{R}^{k_{d}}$ and $\lambda\in[0,1]$, $\phi(\lambda X+(1-\lambda)Y)\leq \lambda\phi(X)+(1-\lambda)\phi(Y)$;
		\item[B3] Correlation: for any $X\in \mathbb{R}^{k_{1}}\times \mathbb{R}^{k_{2}}\times \ldots\times \mathbb{R}^{k_{d}}$, there exists a simple risk statistic $\varrho$ such that $ \big((\varrho \circ \phi)(X_{[k_{1}]}),(\varrho \circ \phi)(X_{[k_{2}]}),\ldots, (\varrho \circ \phi)(X_{[k_{d}]})\big) = \phi(X)$.
	\end{description}
\end{Definition}

\begin{Definition}
	A complex risk statistic is a function $\rho$: $\mathbb{R}^{k_{1}}\times \mathbb{R}^{k_{2}}\times \ldots\times \mathbb{R}^{k_{d}}$ $\rightarrow$ $\mathbb{R}$ that satisfies the following properties,
	\begin{description}
		\item[C1] Monotonicity: for any $X,Y\in \mathbb{R}^{k_{1}}\times \mathbb{R}^{k_{2}}\times \ldots\times \mathbb{R}^{k_{d}}$,  $X\succeq Y$ implies $\rho(X)\geq \rho(Y)$;
		\item[C2] Convexity: for any  $X,Y\in \mathbb{R}^{k_{1}}\times \mathbb{R}^{k_{2}}\times \ldots\times \mathbb{R}^{k_{d}}$ and $\lambda\in[0,1]$, $\rho(\lambda X+(1-\lambda)Y)\leq \lambda\rho(X)+(1-\lambda)\rho(Y)$;
		\item[C3] Statistical convexity: for any $X,Y,Z\in \mathbb{R}^{k_{1}}\times \mathbb{R}^{k_{2}}\times \ldots\times \mathbb{R}^{k_{d}}$, $\lambda\in[0,1]$ and $1\leq i \leq d$, 
		if $ \rho(Z_{[k_{i}]})=\lambda\rho(X_{[k_{i}]})+(1-\lambda)\rho(Y_{[k_{i}]}) $, then $\rho(Z)\leq \lambda\rho(X)+(1-\lambda)\rho(Y)$.
	\end{description}
\end{Definition}

\section{How to measure complex risk }
\label{sec:4}

In this section, we derive a new approach to measure complex risk in intelligent systems. To this end, we show that each complex risk statistic can be decomposed into a simple risk statistic $varrho$ and a clustering function $\phi$. In other words, the measurement of complex risk statistics can be simplified into two steps.

\begin{Theorem}\label{T41}
	A function $\rho$: $\mathbb{R}^{k_{1}}\times \mathbb{R}^{k_{2}}\times \ldots\times \mathbb{R}^{k_{d}}$ $\rightarrow$ $\mathbb{R}$ is a complex risk statistic in the case of there exists a clustering function $\phi:\mathbb{R}^{k_{1}}\times \mathbb{R}^{k_{2}}\times \ldots\times \mathbb{R}^{k_{d}} \rightarrow \mathbb{R}^{d}$ and a simple risk statistic $\varrho:$ $\mathbb{R}^{d} \rightarrow \mathbb{R}$ such that $\rho$ is the composition of $\varrho$ and $\phi$, i.e.
	\begin{equation}\label{41}
	\rho(X) = (\varrho \circ \phi) (X) \qquad \textrm{for all} \quad X\in \mathbb{R}^{k_{1}}\times \mathbb{R}^{k_{2}}\times \ldots\times \mathbb{R}^{k_{d}}.
	\end{equation}
\end{Theorem}

\noindent \textbf{Proof.}
We first derive the ` only if ' part. We suppose that $\rho$ is a complex risk statistic and define a function $\phi$ by
\begin{equation}\label{42}
\phi(X):=\big(\rho(X_{[k_{1}]}),\rho(X_{[k_{2}]}),\ldots, \rho(X_{[k_{d}]})\big)
\end{equation}
for any $X\in \mathbb{R}^{k_{1}}\times \mathbb{R}^{k_{2}}\times \ldots\times \mathbb{R}^{k_{d}}$. 
Since $\rho$ satisfies the convexity $\mathbf{C2}$, it follows
\begin{eqnarray*}
	\phi\big(\lambda X+(1-\lambda)Y\big)&=&\big(\rho(\lambda X_{[k_{1}]}+(1-\lambda)Y_{[k_{1}]}),\ldots, \rho(\lambda X_{[k_{d}]}+(1-\lambda)Y_{[k_{d}]})\big)\\
	&\leq &\lambda\big(\rho(X_{[k_{1}]}),\ldots, \rho(X_{[k_{d}]})\big)+(1-\lambda)\big(\rho(Y_{[k_{1}]}),\ldots, \rho(Y_{[k_{d}]})\big)\\
	&=&\lambda\phi(X)+(1-\lambda)\phi(Y)
\end{eqnarray*}
for any $X,Y\in \mathbb{R}^{k_{1}}\times \mathbb{R}^{k_{2}}\times \ldots\times \mathbb{R}^{k_{d}}$ and $\lambda\in [0,1]$. Thus, $\phi$ satisfies the convexity $\mathbf{B2}$. Similarly, the monotonicity $\mathbf{B1}$ of $\phi$ can also be implied by the monotonicity $\mathbf{C1}$ of $\rho$.  Next, we consider a function $\varrho:$ $\phi(\mathbb{R}^{k_{1}}\times \mathbb{R}^{k_{2}}\times \ldots\times \mathbb{R}^{k_{d}}) \rightarrow \mathbb{R}$ that is defined by
\begin{equation}\label{43}
\varrho(x):=\rho(X)\qquad \textrm{where} \ X\in \mathbb{R}^{k_{1}}\times \mathbb{R}^{k_{2}}\times \ldots\times \mathbb{R}^{k_{d}} \ \textrm{with} \ \phi(X)=x.
\end{equation}
Thus, we immediately know that $\phi$ satisfies the correlation $\mathbf{B3}$, which means  $\phi$ defined above is a clustering function.
Next, we illustrate that the $\varrho$ defined above is a simple risk statistic. Suppose $x,y \in \phi(\mathbb{R}^{k_{1}}\times \mathbb{R}^{k_{2}}\times \ldots\times \mathbb{R}^{k_{d}})$ with $x \geq y$, there exists $X,Y \in \mathbb{R}^{k_{1}}\times \mathbb{R}^{k_{2}}\times \ldots\times \mathbb{R}^{k_{d}}$ such that $\phi(X)=x$, $\phi(Y)=y$. Then, we have
$
\phi(X) \geq \phi(Y), 
$
which means  $X\geq Y$ by the monotonicity of $\phi$. Thus, it follows from the property $\mathbf{C1}$ of $\rho$ that
\begin{displaymath}
\varrho(x)= \rho(X) \geq \rho(Y) = \varrho(y)
\end{displaymath}
which implies $\varrho$ satisfies the monotonicity $\mathbf{A1}$. Let $x,y \in \phi(\mathbb{R}^{k_{1}}\times \mathbb{R}^{k_{2}}\times \ldots\times \mathbb{R}^{k_{d}})$ with  $\phi(X)=x$, $\phi(Y)=y$ for any $X,Y \in \mathbb{R}^{k_{1}}\times \mathbb{R}^{k_{2}}\times \ldots\times \mathbb{R}^{k_{d}}$, which implies $\varrho(x)= \rho(X)$ and $\varrho(y)= \rho(Y)$.
We also consider $z:= \lambda x + (1-\lambda) y$ for any $\lambda\in [0,1]$. Thus, from the definition of $\varrho$, there exists a $Z\in \mathbb{R}^{k_{1}}\times \mathbb{R}^{k_{2}}\times \ldots\times \mathbb{R}^{k_{d}}$ such that 
\[
\varrho(\lambda x + (1-\lambda) y)=\rho(Z)
\]
with $\phi(Z)=\lambda x + (1-\lambda) y$. Hence, from the statistic convexity $\mathbf{C3}$ of $\rho$, we know that
\begin{eqnarray*}
	\varrho(\lambda x + (1-\lambda) y)&=&\rho(Z)\\
	&\leq &\lambda\rho(X)+(1-\lambda)\rho(Y)\\
	&=&\lambda\varrho(x)+(1-\lambda)\varrho(y),
\end{eqnarray*}
which implies the convexity $\mathbf{A2}$ of $\varrho$. 
Thus, $\varrho$ is a simple risk statistic and from (\ref{42}) and (\ref{43}), we have $\rho = \varrho\circ \phi$. Next, we derive the ` if ' part. We suppose that $\phi$ is a clustering function and $\varrho$ is a simple risk statistic. Furthermore, define $\rho = \varrho\circ \phi$. 
Since $\varrho$ and $\phi$ are monotone and convex, it is relatively easy to check that $\rho$ satisfies monotonicity $\mathbf{C1}$ and convexity $\mathbf{C2}$. We now suppose that $X,Y,Z \in \mathbb{R}^{k_{1}}\times \mathbb{R}^{k_{2}}\times \ldots\times \mathbb{R}^{k_{d}}$ which satisfies
\begin{displaymath}
\rho(Z_{[k_{i}]})=\lambda\rho(X_{[k_{i}]})+(1-\lambda)\rho(Y_{[k_{i}]})
\end{displaymath}
for any $\lambda\in [0,1]$. Then, the property $\mathbf{B3}$ of $\phi$ implies
\begin{displaymath}
\phi(Z)= \lambda\phi(X)+(1-\lambda)\phi(Y).
\end{displaymath}
Thus, we have
\begin{eqnarray*}
	\rho(Z)
	&=&\varrho(\lambda\phi(X)+(1-\lambda)\phi(Y))\\
	&\leq&\lambda\rho(X)+(1-\lambda)\rho(Y),
\end{eqnarray*}
which indicates $\rho$ satisfies the property $\mathbf{C3}$. 
Thus, the  $\rho$ defined above is a complex risk statistic.\qed\\

\begin{Remark}
	Theorem~\ref{T41} not only provide a decomposition result for complex risk statistics, but also propose a approach to deal with complex risk especially in large scale integration. Notably, we first use the clustering function $\phi$ to convert the complex system risk into simple, then we quantify the simplified risk by the simple risk statistic.
	Therefore, an engineer who deal with the measurement of complex risk in large scale integration can construct a reasonable complex risk statistic by choosing an appropriate clustering function and an appropriate simple risk statistic. The clustering function should reflect his preferences regarding the uncertainty of large scale integration.
\end{Remark}

In the following section, we derive the dual representation of complex risk statistics with the acceptance sets of  $\phi$ and $\varrho$.

\section{Dual representation}
\label{sec:5}
Before we study the dual representation of complex risk statistics on $\mathbb{R}^{k_{1}}\times \mathbb{R}^{k_{2}}\times \ldots\times \mathbb{R}^{k_{d}}$, the acceptance sets should be defined. Since each complex risk statistic $\rho$ can be decomposed into  a clustering function $\phi$ and a simple risk statistic $\varrho$, we  need only to define the acceptance sets of $\phi$ and $\varrho$, i.e.

\begin{equation}\label{51}
\mathcal{A}_{\varrho}:=\big\{(c,x)\in \mathbb{R}\times \mathbb{R}^{d}: \varrho(x)\leq c \big\}
\end{equation}
and
\begin{equation}\label{52}
\mathcal{A}_{\phi}:=\big\{(y,X)\in \mathbb{R}^{d}\times( \mathbb{R}^{k_{1}}\times \mathbb{R}^{k_{2}}\times \ldots\times \mathbb{R}^{k_{d}}): \phi(X)\leq y \big\}.\\
\end{equation}

We will see later on that these acceptance sets can be used to provide complex risk statistics on $\mathbb{R}^{k_{1}}\times \mathbb{R}^{k_{2}}\times \ldots\times \mathbb{R}^{k_{d}}$. The following properties  are needed in the subsequent study.

\begin{Definition}
	Let $M$ and $N$ be two ordered linear spaces. A set $A \subset M \times N$ satisfies f-monotonicity if $(m,n)\in A$, $q\in N$ and $n\geq q$ imply $(m,q)\in A$.  A set $A \subset M \times N$ satisfies b-monotonicity if $(m,n)\in A$, $p\in M$ and $p\geq m$ imply $(p,n)\in A$.
\end{Definition}

\begin{Proposition}\label{P1}
	We suppose that $\rho = \varrho\circ\phi$ is a complex risk statistic with a clustering function $\phi:\mathbb{R}^{k_{1}}\times \mathbb{R}^{k_{2}}\times \ldots\times \mathbb{R}^{k_{d}} \rightarrow \mathbb{R}^{d}$ and a simple risk statistic $\varrho:$ $\mathbb{R}^{d} \rightarrow \mathbb{R}$. The corresponding acceptance sets $\mathcal{A}_{\varrho}$ and $\mathcal{A}_{\phi}$ are defined by (\ref{51}) and (\ref{52}). Then, $\mathcal{A}_{\phi}$ and $\mathcal{A}_{\varrho}$ are convex sets and they satisfy the f-monotonicity and b-monotonicity.
\end{Proposition}

\noindent \textbf{Proof.}
It is easy to check the above properties from definitions of $\phi$ and $\varrho$. \qed\\

\indent The next proposition provides the primal representation of complex risk statistics on $\mathbb{R}^{k_{1}}\times \mathbb{R}^{k_{2}}\times \ldots\times \mathbb{R}^{k_{d}}$  considering the acceptance sets.

\begin{Proposition}\label{P2}
	We suppose that $\rho = \varrho\circ\phi$ is a complex risk statistic with a clustering function $\phi:\mathbb{R}^{k_{1}}\times \mathbb{R}^{k_{2}}\times \ldots\times \mathbb{R}^{k_{d}} \rightarrow \mathbb{R}^{d}$ and a simple risk statistic $\varrho:$ $\mathbb{R}^{d} \rightarrow \mathbb{R}$. The corresponding acceptance sets $\mathcal{A}_{\varrho}$ and $\mathcal{A}_{\phi}$ are defined by (\ref{51}) and (\ref{52}). Then, for any $X\in \mathbb{R}^{k_{1}}\times \mathbb{R}^{k_{2}}\times \ldots\times \mathbb{R}^{k_{d}}$,
	\begin{equation}\label{53}
	\rho(X) = \inf \big\{c\in \mathbb{R}: (c,x)\in \mathcal{A}_{\varrho}, (x,X)\in \mathcal{A}_{\phi} \big\}
	\end{equation}
	where we set $\inf \emptyset = +\infty$.
\end{Proposition}

\noindent \textbf{Proof.}
Since $\rho = \varrho\circ\phi$, we have
\begin{equation}\label{54}
\rho(X) = \inf \big\{c\in \mathbb{R}:  (\varrho\circ\phi)(X)\leq c \big\}.
\end{equation}
Using the definition of $\mathcal{A}_{\varrho}$, we know that
\begin{equation}\label{55}
\varrho(x) = \inf \big\{c\in \mathbb{R}: (c,x)\in \mathcal{A}_{\varrho}  \big\}
\end{equation}
for any $x\in \mathbb{R}^d$. Then, from (\ref{54}) and (\ref{55}),
\begin{displaymath}
\rho(X) = \inf \big\{c\in \mathbb{R}: (c,\phi(X))\in \mathcal{A}_{\varrho}\big\}.
\end{displaymath}
It is easy to check that
\begin{displaymath}
\big\{c\in \mathbb{R}: (c,\phi(X))\in \mathcal{A}_{\varrho}\big\} = \big\{c\in \mathbb{R}: (c,x)\in \mathcal{A}_{\varrho}, (x,X)\in \mathcal{A}_{\phi} \big\}.
\end{displaymath}
Thus,
\begin{displaymath}
\rho(X) = \inf \big\{c\in \mathbb{R}: (c,x)\in \mathcal{A}_{\varrho}, (x,X)\in \mathcal{A}_{\phi} \big\}.
\end{displaymath}\qed\\

With Proposition~\ref{P2}, we now introduce the main result of this section: the dual representation of complex risk statistics on $\mathbb{R}^{k_{1}}\times \mathbb{R}^{k_{2}}\times \ldots\times \mathbb{R}^{k_{d}}$.

\begin{Theorem}\label{T51}
	We suppose that $\rho = \varrho\circ\phi$ is a complex risk statistic characterized by a continue clustering function $\phi$ and a continue simple risk statistic $\varrho$. Then, for any $X\in \mathbb{R}^{k_{1}}\times \mathbb{R}^{k_{2}}\times \ldots\times \mathbb{R}^{k_{d}}$, $\rho(X)$ has the following form
	\begin{equation}\label{56}
	\rho(X) = \sup_{(\widehat{y},\widehat{X})\in \mathcal{P}} \Big\{\langle\widehat{X},X   \rangle - \alpha (\widehat{y},\widehat{X}) \Big\}
	\end{equation}
	where $\alpha: \mathbb{R}^{d}\times( \mathbb{R}^{k_{1}}\times \mathbb{R}^{k_{2}}\times \ldots\times \mathbb{R}^{k_{d}})$ $\rightarrow$ $\mathbb{R}$ is defined by
	\begin{displaymath}
	\alpha (\widehat{y},\widehat{X}):= \sup_{\substack{(c,x)\in \mathcal{A}_{\varrho}\\(y,Y)\in \mathcal{A}_{\phi}}} \Big\{-c - \langle \widehat{y},(y-x) \rangle + \langle\widehat{X},Y  \rangle \Big\}
	\end{displaymath}
	and
	\begin{displaymath}
	\mathcal{P}:= \big\{ (\widehat{y},\widehat{X})\in \mathbb{R}^{d}\times( \mathbb{R}^{k_{1}}\times \mathbb{R}^{k_{2}}\times \ldots\times \mathbb{R}^{k_{d}}), \alpha (\widehat{y},\widehat{X})< \infty \big\}.
	\end{displaymath}
\end{Theorem}

\noindent \textbf{Proof.}
Using Proposition~\ref{P2}, we have
\begin{displaymath}
\rho(X) = \inf \big\{c\in \mathbb{R}: (c,x)\in \mathcal{A}_{\varrho}, (x,X)\in \mathcal{A}_{\phi} \big\}
\end{displaymath}
for any $X\in\mathbb{R}^{k_{1}}\times \mathbb{R}^{k_{2}}\times \ldots\times \mathbb{R}^{k_{d}}$. Furthermore, we can rewrite this formula as
\begin{equation}\label{57}
\rho(X) = \inf_{(c,x)\in \mathbb{R}\times \mathbb{R}^{d}} \big\{c + I_{\mathcal{A}_{\varrho}}(c,x) + I_{\mathcal{A}_{\phi}}(x,X) \big\}
\end{equation}
where the indicator function of  a set $A \in \mathcal{X}\times \mathcal{Y}$ is defined by
\begin{displaymath}
I_{A}(a,b):=\left\{ \begin{array}{ll}
0, & (a,b)\in \mathcal{X}\times \mathcal{Y}\\
\infty, & \textrm{otherwise.}
\end{array} \right.
\end{displaymath}
From Proposition~\ref{P1}, we know that $\mathcal{A}_{\varrho}$ and $\mathcal{A}_{\phi}$ are convex sets. Thus,
\begin{displaymath}
I_{\mathcal{A}_{\varrho}}^{\prime}(\widehat{c},\widehat{x})= \sup_{(\overline{c},\overline{x})\in \mathcal{A}_{\varrho}} \big\{ \widehat{c} \overline{c} + \langle \widehat{x}, \overline{x}\rangle  \big\},\quad  \widehat{c}\in\mathbb{R}, \widehat{x}\in \mathbb{R}^{d}
\end{displaymath}
and
\begin{displaymath}
I_{\mathcal{A}_{\phi}}^{\prime}(\widehat{y},\widehat{X})= \sup_{(\overline{y},\overline{X})\in \mathcal{A}_{\phi}} \big\{ \widehat{y} \overline{y} + \langle \widehat{X}, \overline{X}\rangle  \big\},\quad  \widehat{y}\in \mathbb{R}^{d}, \widehat{X}\in \mathbb{R}^{k_{1}}\times \mathbb{R}^{k_{2}}\times \ldots\times \mathbb{R}^{k_{d}}.
\end{displaymath}
Next, since $\varrho$ is  continue, it follows that $\mathcal{A}_{\varrho}$ is closed. Thus, by the duality theorem for conjugate functions, we have
\begin{eqnarray*}
	I_{\mathcal{A}_{\varrho}}(c,x) &=& I_{\mathcal{A}_{\varrho}}^{\prime\prime}(c,x)\\&=&\sup_{(\widehat{c},\widehat{x})\in \mathbb{R}\times \mathbb{R}^{d}}\big\{ \widehat{c}c + \langle \widehat{x},x \rangle - I_{\mathcal{A}_{\varrho}}^{\prime}(\widehat{c},\widehat{x}) \big\}\\
	&=& \sup_{(\widehat{c},\widehat{x})\in \mathbb{R}\times \mathbb{R}^{d}}\Big\{ \widehat{c}c + \langle \widehat{x},x \rangle - \sup_{(\overline{c},\overline{x})\in \mathcal{A}_{\varrho}} \big\{ \widehat{c} \overline{c} + \langle \widehat{x}, \overline{x}\rangle  \big\}\Big\}.
\end{eqnarray*}
Similarly, we have
\begin{eqnarray*}
	I_{\mathcal{A}_{\phi}}(x,X) &=& I_{\mathcal{A}_{\phi}}^{\prime\prime}(x,X)\\&=&\sup_{(\widehat{y},\widehat{X})\in \mathbb{R}^{d}\times( \mathbb{R}^{k_{1}}\times \mathbb{R}^{k_{2}}\times \ldots\times \mathbb{R}^{k_{d}})}\big\{  \langle \widehat{y},x\rangle + \langle \widehat{X},X \rangle - I_{\mathcal{A}_{\phi}}^{\prime}(\widehat{y},\widehat{X}) \big\}\\
	&=& \sup_{(\widehat{y},\widehat{X})\in \mathbb{R}^{d}\times( \mathbb{R}^{k_{1}}\times \mathbb{R}^{k_{2}}\times \ldots\times \mathbb{R}^{k_{d}})}\Big\{ \langle \widehat{y},x\rangle + \langle \widehat{X},X \rangle - \sup_{(\overline{y},\overline{X})\in \mathcal{A}_{\phi}} \big\{ \langle \widehat{y},\overline{y}\rangle + \langle \widehat{X}, \overline{X}\rangle  \big\}\Big\}.
\end{eqnarray*}
Thus, we know that
\begin{eqnarray*}
	\rho(X)&=& \inf_{(c,x)\in \mathbb{R}\times \mathbb{R}^d} \big\{c + I_{\mathcal{A}_{\varrho}}(c,x) + I_{\mathcal{A}_{\phi}}(x,X) \big\}\\
	&=& \inf_{(c,x)\in \mathbb{R}\times \mathbb{R}^d} \sup_{\substack {(\widehat{c},\widehat{x})\in \mathbb{R}\times \mathbb{R}^{d}\\(\widehat{y},\widehat{X})\in \mathbb{R}^{d}\times( \mathbb{R}^{k_{1}}\times \mathbb{R}^{k_{2}}\times \ldots\times \mathbb{R}^{k_{d}})}}\Big\{ c(1+\widehat{c}) +  \langle \widehat{x}+ \widehat{y}, x\rangle + \langle \widehat{X},X \rangle - I_{\mathcal{A}_{\varrho}}^{\prime}(\widehat{c},\widehat{x})-\\ && I_{\mathcal{A}_{\phi}}^{\prime}(\widehat{y},\widehat{X})  \Big\}.
\end{eqnarray*}
From the continuity of $\varrho$ and the continuity of $\phi$, we can interchange the supremum and the infimum above, i.e.
\begin{eqnarray*}
	\rho(X) &=& \sup_{\substack {(\widehat{c},\widehat{x})\in \mathbb{R}\times \mathbb{R}^{d}\\(\widehat{y},\widehat{X})\in \mathbb{R}^{d}\times( \mathbb{R}^{k_{1}}\times \mathbb{R}^{k_{2}}\times \ldots\times \mathbb{R}^{k_{d}})}}\inf_{(c,x)\in \mathbb{R}\times \mathbb{R}^d}\Big\{ c(1+\widehat{c}) +  \langle \widehat{x}+ \widehat{y}, x\rangle + \langle \widehat{X},X \rangle - I_{\mathcal{A}_{\varrho}}^{\prime}(\widehat{c},\widehat{x})- I_{\mathcal{A}_{\phi}}^{\prime}(\widehat{y},\widehat{X})  \Big\}\\
	&=& \sup_{(\widehat{y},\widehat{X})\in \mathbb{R}^{d}\times( \mathbb{R}^{k_{1}}\times \mathbb{R}^{k_{2}}\times \ldots\times \mathbb{R}^{k_{d}})} \Big\{  \langle \widehat{X},X \rangle - \sup_{\substack {(\overline{c},\overline{x})\in \mathcal{A}_{\varrho}\\(\overline{y},\overline{X})\in \mathcal{A}_{\phi}}} \big\{  -\overline{c}-  \langle \widehat{y}, \overline{y}-\overline{x}\rangle +   \langle \widehat{X},\overline{X} \rangle \big\}  \Big\}.
\end{eqnarray*}
With $ \alpha (\widehat{y},\widehat{X})$ is defined by
\begin{eqnarray*}
	\alpha (\widehat{y},\widehat{X}):&=&  \sup_{\substack {(\overline{c},\overline{x})\in \mathcal{A}_{\varrho}\\(\overline{y},\overline{X})\in \mathcal{A}_{\phi}}} \big\{  -\overline{c}-  \langle \widehat{y}, \overline{y}-\overline{x}\rangle +   \langle \widehat{X},\overline{X} \rangle \big\}\\
	&=&\sup_{\substack{(c,x)\in \mathcal{A}_{\varrho}\\(y,Y)\in \mathcal{A}_{\phi}}} \Big\{-c - \langle \widehat{y},(y-x) \rangle + \langle\widehat{X},Y  \rangle \Big\}
\end{eqnarray*}
and
\begin{displaymath}
\mathcal{P}:= \big\{ (\widehat{y},\widehat{X})\in \mathbb{R}^{d}\times( \mathbb{R}^{k_{1}}\times \mathbb{R}^{k_{2}}\times \ldots\times \mathbb{R}^{k_{d}}), \alpha (\widehat{y},\widehat{X})< \infty \big\},
\end{displaymath}
it immediately follows that
\begin{displaymath}
\rho(X) = \sup_{(\widehat{y},\widehat{X})\in \mathcal{P}} \Big\{\langle\widehat{X},X   \rangle - \alpha (\widehat{y},\widehat{X}) \Big\}.
\end{displaymath}\qed\\

\begin{Remark}
	Note that, the proof of Theorem~\ref{T51} above utilized the primal representation of complex risk statisticss in Proposition~\ref{P2}, which indicates that the acceptance sets $\mathcal{A}_{\varrho}$ and $\mathcal{A}_{\phi}$ play a vital role. Thus, the  dual representation of complex risk statistics $\rho$  still dependent on the clustering function $\phi$ and the simple risk statistic $\varrho$.
\end{Remark}

\section{Conclusions} 
In this paper, we derive a new class of risk statistics in intelligent systems, especially in deep neural networks, named complex risk statistics. Our results illustrate that an engineer who deal with the measurement of complex risk in intelligent systems can construct a reasonable complex risk statistic by choosing an appropriate clustering function and an appropriate simple risk statistic.

%\section*{Author Contributions}
% Conceptualization, F.Sun and Y.Dong; Formal analysis, F.Sun and Y.Dong; Writing-original draft preparation, F.Sun and Y.Dong; Writing-review and editing, F.Sun and Y.Dong; Funding acquisition, F.Sun.

%\section*{Funding Statement}
%This work is supported by Young Innovative Talents Project of Guangdong Province  (2019KQNCX156).

%\section*{Data Availability Statement}

%No data, code were generated or used during the study.

%\section*{Conflict of Interest Statement}

%The authors declare that the research was conducted in the absence of any commercial or financial relationships that could be construed as a potential conflict of interest.

%\section*{Acknowledgements}
%This manuscript has been released as a pre-print at [http://export.arxiv.org/pdf/2003.09255], \cite{16}.


\begin{thebibliography}{99}
	{
		
		
		\bibitem{1}   V.V. Acharya,  L.H. Pedersen, T. Philippon, M. Richardson, (2012). easuring systemic risk. CEPR Discussion Paper 8824,  http://www.cepr.org/pubs/dps/DP8824.asp
		\bibitem{3} Y. Armenti, S. Crepey, S. Drapeau, A. Papapantoleon, (2015). Multivariate shortfall risk allocation and systemic risk. arXiv: 1507.05351 
		
		
		\bibitem{2} S. Ahmed, D. Filipovi\'{c}, and G. Svindland, (2008). A note on natural risk statistics, Oper. Res. Lett. 36, 662-664.
		
		
		\bibitem{4} P. Artzner, F. Dellbaen, J.M. Eber, D. Heath, (1997). Thinking coherently. Risk. 10, 68-71. 
		\bibitem{5} P. Artzner, F. Dellbaen, J.M. Eber, D. Heath, (1999). Coherent measures of risk. Math. Finance. 9(3), 203-228. 
 
		
		\bibitem{6} F. Biagini, J.P. Fouque, M. Frittelli, (2015). A unified approach to systemic risk measures via acceptance sets. arXiv: 1503.06354.
		
		\bibitem{7} M.K. Brunnermeier, P. Cheridito, (2014). Measuring and allocating systemic risk, http://ssrn.com/abstract=2372472
		
		\bibitem{8} C. Chen, G. Iyengar,  C.C. Moallemi, (2013). An axiomatic approach to systemic risk. Manage. Sci. 59(6), 1373-1388.
		
		\bibitem{15} X. Deng, F. Sun, (2020). Regulator-based risk statistics for portfolios, Discrete. Dyn. Nat. Soc. https://doi.org/10.1155/2020/7015267
		
		\bibitem{16} X. Deng, F. Sun, (2020). Systemic risk statistics with scenario analysis. arXiv: 2003.09255 
		
		\bibitem{11} Z. Feinstein, B. Rudloff, S. Weber, (2015). Measures of systemic risk, arXiv: 1502.07961.
		
		\bibitem{13} C. Gauthier, M. Lehar, M. Souissi, (2012). Macroprudential capital requirements and systemic risk. J. Financ Intermed. 21(4), 594-618 .
		
		
		
		\bibitem{18} C.C. Heyde, S.G. Kou, and X.H. Peng, (2007). What is a good external risk measure: Bridging the gaps between robustness, subadditivity, and insurance risk measures, Working paper, Columbia University. 
		
		\bibitem{20} S.G. Kou, X.H. Peng, and C.C. Heyde, (2013). External risk measures and basel accords, Math. Oper. Res. 38, 393-417.
		
		\bibitem{21} F. Sun,  Y.H. Chen,  Y.J. Hu,  (2018). Set-valued loss-based risk measures, Positivity, 22(3), 859-871.
		\bibitem{22} F. Sun, Y.J. Hu, (2019). Set-valued cash sub-additive risk measures, Probab. Engrg. Inform. Sci. 33(2), 241-257.
		
		\bibitem{16}  F. Sun, Y. Dong, (2020). Systemic risk statistics with scenario analysis. arXiv: 2003.09255 
		
		\bibitem{23} N. Tarashev, C. Borio, K. Tsatsaronis, (2010). Attributing systemic risk to individual institutions. Working Paper No. 308. Bank for International Settlements, Basel, http://www.bis.org/publ/work308.pdf
		
		
		
		\bibitem{25} D.J. Tian, L. Jiang, (2015). Quasiconvex risk statistics with scenario analysis, Math. Financ. Econ. 9. 111-121.
		\bibitem{26} D.J. Tian, X.L. Suo, (2012). A note on convex risk statisitc, Oper. Res. Lett. 40. 551-553.
		
		
		
	}
\end{thebibliography}
\end{document}